\documentclass[prb,twocolumn,superscriptaddress]{revtex4}

\usepackage{graphicx}

\begin{document}

\title{Calculations of spin-disorder resistivity from first principles}

\author{A. L. Wysocki}
\email{awysocki@bigred.unl.edu} \affiliation{Department of Physics
and Astronomy and Nebraska Center for Materials and Nanoscience,
University of Nebraska--Lincoln, Lincoln, Nebraska 68588, USA}
\author{K. D. Belashchenko}
\affiliation{Department of Physics and Astronomy and Nebraska Center
for Materials and Nanoscience, University of Nebraska--Lincoln,
Lincoln, Nebraska 68588, USA}
\author{J. P. Velev}
\affiliation{Department of Physics and Astronomy and Nebraska Center
for Materials and Nanoscience, University of Nebraska--Lincoln,
Lincoln, Nebraska 68588, USA}
\author{M. van Schilfgaarde}
\affiliation{Department of Chemical and Materials Engineering,
Arizona State University, Tempe, Arizona 85287, USA}

\date{\today}

\begin{abstract}
Spin-disorder resistivity of Fe and Ni is studied using the
noncollinear density functional theory. The Landauer conductance is
averaged over random disorder configurations and fitted to Ohm's
law. The distribution function is approximated by the mean-field
theory. The dependence of spin-disorder resistivity on magnetization
in Fe is found to be in excellent agreement with the results for the
isotropic $s$-$d$ model. In the fully disordered state,
spin-disorder resistivity for Fe is close to experiment, while for
fcc Ni it exceeds the experimental value by a factor of 2.3. This
result indicates strong magnetic short-range order in Ni at the
Curie temperature.
\end{abstract}

\maketitle


The effects of thermal spin disorder on the transport properties of
ferromagnets are of interest for both fundamental and practical
reasons. First, studies of spin disorder resistivity may provide
important information on magnetic short-range order in ferromagnets,
which is an intrinsic thermodynamic feature of spin disorder.
Second, spin disorder intermixes the spin channels and thereby
introduces finite spin diffusion length in ferromagnets. Together
with scattering on interface spin disorder and spin-orbit
scattering, this effect degrades the magnetoresistance of magnetic
tunnel junctions and spin valves, which are used in magnetic field
sensors and magnetic random-access memories.

A large amount of experimental data is available on the electric
resistivity of ferromagnetic metals, most of which was collected
many years ago. \cite{Coles}  In these metals the resistivity has an
``anomalous'' contribution which contains signatures of magnetic
phase transitions. This magnetic contribution $\rho_\mathrm{mag}$
may be extracted from the measured resistivity assuming that the
Matthiessen's rule is valid, and that the anomalous contribution is
temperature-independent well above the Curie
temperature.\cite{Weiss}

The anomalous resistivity $\rho_\mathrm{mag}$ may be attributed to
spin-disorder scattering, which was studied using the $s$-$d$ model
Hamiltonian by many authors.\cite{Kasuya,dGF,Mannari,Vonsovskii}
This approach assumes that $3d$ electrons in transition metals are
localized at atomic sites, while the conductivity is due to mobile
$4s$ electrons forming an itinerant band and coupled to the $d$
electrons by exchange interaction. At finite temperatures the
directions of the $d$-electron spins fluctuate, and the conducting
$s$ electrons scatter from the inhomogeneous exchange potential.

As the temperature is increased toward the Curie temperature $T_c$,
the spins become more disordered, and $\rho_\mathrm{mag}$ quickly
increases, sometimes surpassing the phonon contribution. At elevated
temperatures the $s$-$d$ model calculations of $\rho_\mathrm{mag}$
were done using the mean-field approximation for magnetic
thermodynamics. In this approximation, the spins are completely
disordered above $T_c$, and hence $\rho_\mathrm{mag}$ is constant.
In the Born approximation below $T_c$ it declines as
\begin{equation}
\rho_\mathrm{mag}(T)=\rho_\mathrm{mag}(T_c)\left[1-M^2(T)/M^2(0)\right],
\label{rhovsT}
\end{equation}
where $M(T)$ is the magnetization at temperature
$T$.\cite{Kasuya,dGF,Mannari} In general, $\rho_\mathrm{mag}$ is
sensitive to magnetic short-range, rather than long-range
order.\cite{Fisher} Quantitative comparison of the calculations of
$\rho_\mathrm{mag}(T)$ with experiment can thus provide valuable
information on magnetic short-range order, which may be very strong
in itinerant ferromagnets like Ni.\cite{Antropov}

Unfortunately, since the $s$-$d$ model assumes an isotropic
conduction band, it can not predict the magnitude of
$\rho_\mathrm{mag}(T_c)$ of specific materials. Moreover, treating
the $3d$ electrons as localized appears to be an inadequate
approximation for computing the spin-disorder resistivity, because
the dominant contribution comes from interband ($s$-$d$), rather
than intraband ($s$-$s$) scattering transitions. This conclusion was
made by Goodings \cite{Goodings} who solved the Boltzmann equation
for the isotropic two-band model. Although this model is not
applicable to transition metals with complex Fermi surfaces, it
clearly highlights the sensitivity of the spin-disorder resistivity
to the band structure. A parameter-free approach based on the
density functional theory is clearly needed for quantitative
comparison with experiment.

In this paper we calculate $\rho_\mathrm{mag}$ for iron and nickel
from first principles using the Kubo-Landauer formalism implemented
in the tight-binding linear muffin-tin orbital (TB-LMTO) method
within the atomic sphere approximation, which was extended to
non-collinear systems. Our treatment of spin disorder is based on
the adiabatic approximation, which assumes that the low-lying
excitations of the magnet are well represented by the states where
the magnetic moments at different atomic sites are rigidly rotated
off of the magnetization axis.\cite{Gyorffy} The state of the system
is assumed to be the ground state constrained by these rotations.
For a given noncollinear spin configuration, one can thus construct
a TB-LMTO Hamiltonian or the Green's function matrix (which are no
more diagonal in spin space).\cite{noncol} We used the local spin
density approximation and frozen atomic potentials taken from the
ferromagnetic ground state. The basis set included the states up to
$l_\mathrm{max}=2$ for both Fe and Ni ($s$, $p$, and $d$ orbitals).

\begin{figure}
\includegraphics*[width=0.45\textwidth]{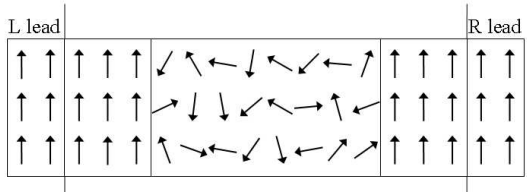}
\caption{The schematic picture of the system used in the
calculations. Vertical lines indicate the embedding
planes.}\label{fig1}
\end{figure}


The conductance was calculated using the principal-layer Green's
function technique and the Kubo-Landauer
formalism.\cite{Turek,Kudrnovsky} We consider a ferromagnetic layer
FM(D) with spin disorder placed between two magnetically ordered
leads FM(O) made of the same material, as seen in Fig.~\ref{fig1}.
As usual in the principal-layer technique, the ``active region''
consists of the FM(D) layer plus a few monolayers of FM(O) on each
side.

Because we use frozen atomic potentials to make calculations
practical, care needs to be taken to ensure local charge neutrality.
Indeed, one can view FM(D) and FM(O) as different materials which
have different Fermi levels that must normally be matched by the
contact potential. Frozen atomic potentials do not allow this to
happen, and therefore we introduced a compensating constant
potential shift in the disordered region which mimics the contact
potential. This shift was found so that the charge in the central
part of FM(D) averaged over disorder realizations is zero. Note that
no matter how the FM(O)/FM(D) interfaces are treated
(self-consistently or not), they add contact resistances to the
circuit. However, since the \emph{resistivity} of the FM(D) material
is extracted from the dependence of the resistance on the FM(D)
thickness in the Ohmic limit, it does not depend on the interfaces,
and therefore our crude treatment of the interfaces does not
introduce any error.

\begin{figure}
\includegraphics*[width=0.45\textwidth]{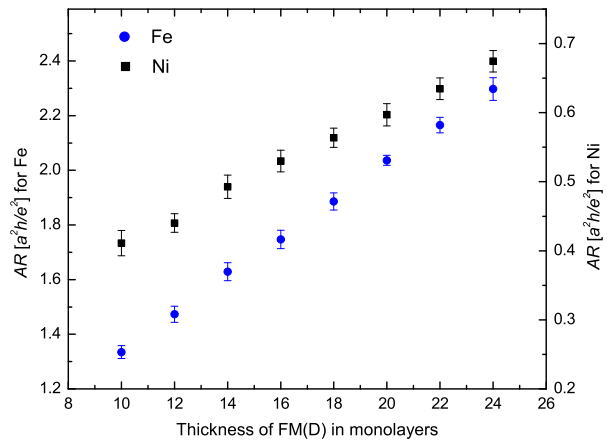}
\caption{The area-resistance product $AR$ of the FM(O)/FM(D)/FM(O)
systems as a function of the thickness of the FM(D) layer for bcc Fe
and fcc Ni in the paramagnetic state. The lateral cell sizes were
$4\times4$ for Fe and $2\times2$ for Ni, with edges along the [100]
directions. The errorbars denote standard deviations of $AR$ due to
the finite number of spin disorder configurations.} \label{fig2}
\end{figure}

First we considered the paramagnetic state with no magnetic
short-range order (completely uncorrelated spins with uniform
angular distribution function). For the given thickness of the FM(D)
layer, the resistance of the system was averaged over several
disorder configurations (typically 12). As seen in Fig.~\ref{fig2},
almost perfect Ohmic behavior was found for Fe down to the smallest
thicknesses for the $4\times4$ lateral cell (16 atoms per layer).
The convergence with respect to the lateral cell size was also
checked. We found that for paramagnetic Fe $\rho_\mathrm{mag}$ is
quite insensitive to this size, as seen in Table ~\ref{tab1}. For Ni
the thickness dependence of $\rho_\mathrm{mag}$ is also fairly close
to linear for the $2\times2$ lateral cell (8 atoms per layer).

The calculated $\rho_\mathrm{mag}$ for Fe (see Table \ref{tab1}) is
about $20\%$ greater than the experimental estimate.\cite{Weiss}
This result indicates that magnetic short-range order in Fe at $T_c$
is weak. However, for the similar model of fcc Ni
$\rho_\mathrm{mag}$ is nearly 2.3 times greater than the
experimental estimate. This large difference indicates either very
strong magnetic short-range order, or a strongly reduced local
moment at $T\sim T_c$.

\begin{table}
\caption{Spin-disorder resistivity $\rho_{mag}$ in
$\mu\Omega\cdot$cm for paramagnetic bcc Fe and fcc Ni. The
calculated values are given for different lateral cell sizes with
edges along the [100] directions. The experimental values are the
high-temperature asymptotes taken from Ref.
\onlinecite{Weiss}.}\label{tab1}
\begin{ruledtabular}
\begin{tabular}{ccccc}
Lateral cell size & $2\times2$ & $3\times3$ & $4\times4$ & Experiment\\
\hline
Fe  & 105.5 & 100.1 & 101.9 & 80\\
Ni  & 34 & & & 15\\
\end{tabular}
\end{ruledtabular}
\end{table}

Further, we calculated $\rho_\mathrm{mag}$ for ferromagnetic Fe in a
range of temperatures. The spin configurations in the disordered Fe
layer were generated using the mean-field theory, i.e. with an
uncorrelated distribution function given by

\newcommand{\bmu}{\mbox{\boldmath$\mu$}}
\begin{equation}
p(\theta) \propto e^{-\mathbf{H}_{\mathrm{eff}}\cdot \bmu
/T}\,,\quad H_{\mathrm{eff}}(T)=\frac{3M(T)T_c}{\mu M(0)}
\end{equation}
where $\theta$ is the angle between the spin and the magnetization
axis, $\mathbf{\mu}$ is the magnetic moment at each site, $M(T)$ is
the magnetization at temperature $T$ in the mean-field
approximation, and $\mathbf{H}_{\mathrm{eff}}$ is the effective
Weiss field.

We used $4\times4$ lateral cells for all temperatures and
thicknesses. The Ohmic limit (linear dependence of the conductance
on length) was achieved for all temperatures down to about $T_c/3$.
This behavior agrees with what one would expect from simple
mean-free path considerations. If we estimate the mean-free path as
$l = \frac{3}{4}AR_{\mathrm{bal}}/\rho$, where $AR_{\mathrm{bal}}$
is the ballistic area-resistance product (this formula is exact for
the free-electron model), we find that $l$ does not exceed the
lateral cell size in this temperature range. Therefore, quantum
fluctuations of the conductance are small, and the transport is
diffusive. Another indication of the Ohmic behavior comes from the
distribution of the current over the spin channels. The conductance
is a sum of four partial conductances, $G_{\uparrow\uparrow}$,
$G_{\downarrow\downarrow}$, $G_{\uparrow\downarrow}$,
$G_{\downarrow\uparrow}$. Spin-conserving and spin-flip scattering
have similar rates in our spin-disorder problem, and therefore the
electrons lose memory of their spin over their mean-free path.
Therefore, in the Ohmic limit the partial conductances must be
proportional to the number of channels in the left and right leads
for the corresponding spin channels: $G_{\sigma\sigma^\prime}\propto
M^L_\sigma M^R_{\sigma^\prime}$. This behavior was indeed observed
down to $T\sim T_c/3$.

\begin{figure}
\begin{center}
\includegraphics[width=0.4\textwidth]{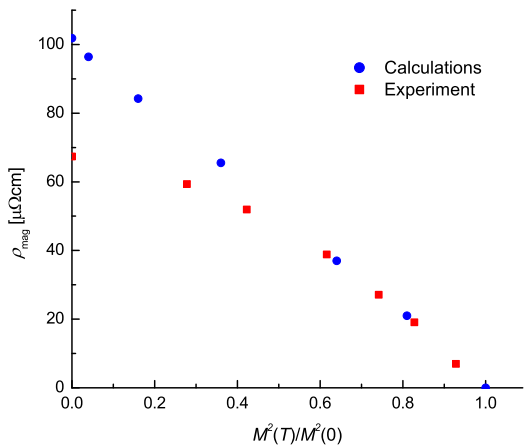}
\end{center} \caption{Spin-disorder resistivity of Fe plotted against
$M^2(T)$. The experimental data for $\rho_\mathrm{mag}(T)$ were
taken from Ref. \onlinecite{Weiss}, and those for $M(T)$ from Ref.
\onlinecite{Crangle}.} \label{fig3}
\end{figure}

The results of $\rho_\mathrm{mag}(T)$ calculations for Fe are
plotted in Fig. \ref{fig3} versus $M^2(T)$ along with the
experimental data taken from Ref. \onlinecite{Weiss}. Surprisingly,
the calculated results, which are based on first-principles
electronic structure and do not involve perturbation theory, show
that the temperature dependence agrees with Eq. (\ref{rhovsT}) with
a rather high accuracy. This striking agreement suggests that the
temperature dependence of $\rho_\mathrm{mag}$ is quite insensitive
to the shape of the Fermi surface.

In the low-temperature region (large $M$) the magnetic excitations
are in reality dominated by spin waves, and the mean-field theory is
inapplicable. Therefore, the fact that the theoretical and
experimental curves in Fig. \ref{fig3} have the same slope in this
region is probably accidental (the extraction of $\rho_\mathrm{mag}$
using Matthiessen's rule is also unreliable here). In the
high-temperature region the comparison is more meaningful, and here
we see that the calculated results are quite close to experiment at
$M^2(T)/M^2(0)\sim0.5$ and somewhat diverge as the temperature
approaches $T_c$. There may be several reasons for this behavior,
including the dependence of the local moment on temperature which we
neglected, and also the presence of magnetic short-range order. The
resistivity is sensitive to short-range, rather than long-range
order \cite{Fisher}, and the short-range order depends less strongly
on the temperature. Therefore, the deviation of the theoretical
curve from experiment close to $T_c$ is expected. Quantitative
analysis of these effects deserves further investigation.

In conclusion, we calculated the spin-disorder contribution to
resistivity for iron and nickel from first-principles. Comparison
with experimental data suggests that magnetic short-range order is
weak in Fe and strong in Ni. The analysis of spin-disorder
resistivity can thus provide important information on magnetic
short-range order in magnetic metals. We also studied the
temperature dependence of spin-disorder resistivity in Fe assuming
mean-field thermodynamics. The resistivity depends linearly on
$M^2(T)/M^2(0)$ in agreement with the predictions of the isotropic
$s$-$d$ model.

We are grateful to V. P. Antropov and E. Y. Tsymbal for useful
discussions. This work was supported by the Nebraska Research
Initiative and NSF MRSEC. It was completed utilizing the Research
Computing Facility of the University of Nebraska--Lincoln.

\end{document}